\documentclass[traditabstract]{aa} 
\usepackage{graphicx}
\usepackage{txfonts}
\usepackage{amssymb,natbib,defs}
\begin{document}
   \title{Disk, merger, or outflow ? Molecular gas kinematics in two
powerful obscured QSOs at $z$$\geq$3.4}


   \author{M. Polletta
          \inst{1}\fnmsep\thanks{Based on observations carried out with the
          IRAM Plateau de Bure Interferometer. IRAM is supported by
          INSU/CNRS (France), MPG (Germany) and IGN (Spain).}
          \and
          N.~P.~H. Nesvadba\inst{2}
          \and
          R. Neri\inst{3}
          \and
          A. Omont\inst{4}
          \and
          S. Berta\inst{5}
          \and
          J. Bergeron\inst{4}
          }
   \institute{INAF -- IASF Milano, via E. Bassini, 20133 Milan, Italy\\
              \email{polletta@iasf-milano.inaf.it}
         \and
             Institut d'Astrophysique Spatiale, Universit\'e de Paris XI, 91405 Orsay Cedex, France\\
         \and 
             Institute de Radioastronomie Millimetrique, 300 rue de la Piscine, 38406 St. Martin d'Heres, France\\
         \and
             Institut d'Astrophysique de Paris, CNRS \& Universit\'e Pierre et Marie Curie, 98bis, bd. Arago, 75014 Paris, France\\
         \and
             Max-Planck-Institut f\"ur extraterrestrische Physik, Postfach 1312, 85741 Garching, Germany
             }

   \date{Submitted on January 6$^{th}$, 2011. Accepted on June 7$^{th}$, 2011.}

  \abstract{We report on the detection of bright CO(4-3) line emission in
two powerful, obscured quasars discovered in the SWIRE survey, SW022513 and
SW022550 at z$\ge$3.4. We analyze the line strength and profile to determine
the gas mass, dynamical mass and the gas dynamics for both galaxies. In
SW022513 we may have found the first evidence for a molecular, AGN-driven
wind in the early Universe. The line profile in SW022513 is broad
(FWHM\,=\,1000\,\kms) and blueshifted by $-$200\,\kms\ relative to systemic
(where the systemic velocity is estimated from the narrow components of
ionized gas lines, as is commonly done for AGN at low and high redshifts).
SW022550 has a more regular, double-peaked profile, which is marginally
spatially resolved in our data, consistent with either a merger or an
extended disk.

The molecular gas masses, $4\times 10^{10}$\,\msun, are large and account
for $<$30\% of the stellar mass, making these obscured QSOs as gas rich as
other powerful CO emitting galaxies at high redshift, i.e., submillimeter
galaxies. Our sources exhibit relatively lower star-formation
efficiencies compared to other dusty, powerful starburst galaxies at high
redshift. We speculate that this could be a consequence of the AGN
perturbing the molecular gas.}


   \maketitle
%

\section{Introduction}

The most accredited galaxy evolution models postulate that the progenitors
of massive galaxies evolve through a phase of intense star-formation
activity at high redshift, accompanied by accretion onto a central
supermassive black
hole~\citep[e.g.][]{sanders88,granato01,granato04,dimatteo05}. Star
formation and accretion are both fueled by cold gas, and are therefore
intricately linked. Several cosmological models now postulate that the phase
of growth of black hole and host is terminated when the accreting black hole
is able to heat and expel the ambient gas
~\citep[e.g.,][]{sanders88,granato01,granato04,dimatteo05,hopkins05b,churazov05,merloni08,fanidakis11}. 
Given the central role of cold gas in these processes, molecular emission
line studies should be ideal to test and refine these
scenarios~\citep[e.g.,][]{narayanan09}.

Owing to their brightness, CO emission lines are commonly used to trace
the immense reservoirs of molecular gas in high-redshift galaxies, providing
valuable constraints on the gas content, gas heating and cooling processes,
and on the evolutionary and dynamical state of these systems. CO line
profiles, and in particular the molecular gas kinematics observed with
interferometers at high spatial resolution, allow to differentiate between,
e.g., rotating disks and mergers. Under the assumption that the gas is
driven by gravity and is approximately virialized, CO line profiles and
velocity gradients can be used to estimate dynamical masses. In recent
years, CO observations have led to a plethora of new constraints, e.g., on
star-formation efficiencies in different types of high-$z$
galaxies~\citep{greve05,solomon05,tacconi08,coppin08b,iono09,daddi10b,riechers11},
the regulation of star-formation in the most rapidly growing
galaxies~\citep{beelen04,greve05,daddi08,schinnerer08,nesvadba09a,genzel10},
or the co-evolution of black holes and their host
galaxies~\citep{peng06b,maiolino07,alexander08,riechers09a,wang10}.

Molecular gas studies have been carried out in starburst-dominated galaxies,
e.g., submillimeter galaxies (SMGs), type I QSOs, radio galaxies, and even
normal star-forming galaxies over large redshift
ranges~\citep[e.g.,][]{omont96,neri03,solomon05,weiss05,greve05,carilli06,tacconi08,coppin08b,nesvadba09a,riechers09a,riechers09b,daddi10b,bothwell10,riechers11}. 
By comparison, only few obscured (type II) quasars have been targeted in CO
millimeter studies~\citep{aravena08,martinez09,yan10}. This is unfortunate,
because obscured quasars could represent the short, but critical transition
phase between starburst and type I (unobscured) QSO
activity~\citep[e.g.,][]{sanders88,hopkins06b}, but has also been
unavoidable so far, because identifying these objects and obtaining
spectroscopic redshifts remains a challenge.

The {\it Spitzer Space Telescope}~\citep{werner04} unveiled a significant
number of obscured quasars owing to their bright mid-infrared (MIR)
emission. However, only about 20-25\% of these galaxies are also bright at
millimeter wavelengths~\citep{lutz05,polletta09,martinez09}, and only few of
those have accurate spectroscopic redshifts, which enable detailed studies
of their millimeter CO emission~\citep{sajina08,yan10}. Those that have been
detected in CO have large molecular gas masses and a wide range of CO line
profiles~\citep{aravena08,martinez09,yan10}. Their overall properties are
similar to SMGs, suggesting they may be mergers~\citep{dasyra08,yan10} with
intense, dusty star formation. Obscured quasars are expected to be close to the
`blowout' phase when the AGN heats and expels the remaining cold ISM,
terminating star formation and accretion activity. However, no signature of such winds has
so far been found, although broad CO line wings have now been identified in
a growing number of nearby AGN, illustrating that molecular outflows are
indeed possible~\citep{feruglio10,alatalo11}.

Here we present CO millimeter observations of two obscured QSOs
associated with massive, intensely star-forming galaxies. These
systems are amongst the most luminous obscured QSOs at high-$z$
currently known. They are thus ideal candidates to investigate if
obscured QSOs are indeed close to complete their phase of active
growth, and in particular, if this phase is terminated by feedback
from their AGN. We will argue that this could potentially be the case
in one of our targets. Throughout the paper, we adopt a
$\Lambda_{CDM}$ cosmology with H$_0$=71 km s$^{-1}$ Mpc$^{-1}$,
$\Omega_{\Lambda}$=0.73, and $\Omega_{M}$=0.37~\citep{spergel03}.

\begin{table}[h!]
\caption{\label{co_params}Redshifts, masses, luminosities, and CO parameters.}
\centering
\begin{tabular}{lrr}
\hline\hline
Parameter&SW022550&SW022513\\
\hline
 $z_{UV}$                                         &    3.867$\pm$0.009 &            \nodata \\
 $z_{opt}$                                        &    3.876$\pm$0.001 &    3.427$\pm$0.002 \\
 F$_{1.2mm}$ [mJy]                                &      4.70$\pm$0.77 &      5.53$\pm$0.72 \\
 M$_{stellar}$\tablefootmark{a} [10$^{10}$\msun]  &             16--80 &             20--40 \\
 Log(L(IR)\tablefootmark{a}) [\lsun]              &         12.5--13.3 &         12.5--13.2 \\
 SFR\tablefootmark{a} [\msun/yr]                  &          500--3000 &          500--3000 \\
 $z_{CO}$                                         &  3.8719$\pm$0.0008 &  3.4220$\pm$0.0007 \\
 $z_{CO,1}$                                       &  3.8686$\pm$0.0007 &  3.4208$\pm$0.0013 \\
 $z_{CO,2}$                                       &  3.8761$\pm$0.0012 &  3.4270$\pm$0.0007 \\
 $\nu$\tablefootmark{b}  [GHz]                    &            94.6330 &           104.2598 \\
 $\nu_1$\tablefootmark{b} [GHz]                   &            94.6978 &           104.2885 \\
 $\nu_2$\tablefootmark{b} [GHz]                   &            94.5504 &           104.1432 \\
 FWHM\tablefootmark{b} [\kms]                     &        800$\pm$120 &       1020$\pm$110 \\
 FWHM$_1$\tablefootmark{b} [\kms]                 &         290$\pm$90 &        950$\pm$150 \\
 FWHM$_2$\tablefootmark{b} [\kms]                 &        420$\pm$180 &        200$\pm$150 \\
 $\Delta$v\tablefootmark{c} [\kms]                &         460$\pm$90 &        420$\pm$100 \\
 I$_{CO}$ [Jy\,\kms]                              &      1.46$\pm$0.28 &      2.62$\pm$0.36 \\
 I$_{CO,1}$ [Jy\,\kms]\tablefootmark{d}           &      0.68$\pm$0.11 &      2.25$\pm$0.28 \\
 I$_{CO,2}$ [Jy\,\kms]\tablefootmark{d}           &      0.70$\pm$0.23 &      0.28$\pm$0.21 \\
 L$_{CO}$\tablefootmark{e} [10$^{8}$\,\lsun]      &      1.77$\pm$0.32 &      2.61$\pm$0.36 \\
 L$^{\prime}_{CO}$\tablefootmark{f} [10$^{10}$\,K\,\kms\,pc$^2$]& 5.7$\pm$1.0 & 8.3$\pm$1.2 \\           
 M$_{gas}$\tablefootmark{g} [10$^{10}$\,\msun]    &     4.53$\pm$0.82  &      6.66$\pm$0.92 \\
 M$_{gas}$/M$_{stellar}$                          &        0.06--0.29  &        0.17--0.33  \\
 M$_{dyn}^{disk}\times sin^2(i)$ [10$^{10}$\,\msun]  &      40$\pm$20  &        2--34       \\
 M$_{dyn}^{merger}\times sin^2(i)$ [10$^{10}$\,\msun]&      20$\pm$10  &        9--35       \\
 SFE\tablefootmark{h} [\lsun/(K\,\kms\,pc$^2$)]   &            35--194 &           24--108  \\
\hline
\end{tabular}
\tablefoot{All CO parameters refer to the $J$=(4--3) transition.\\
\tablefoottext{a}{L(IR) is the 8--1000\,$\mu$m luminosity in \lsun,
M$_{stellar}$ the stellar mass, and SFR the star-formation
rate~\citep[see~\S~\ref{src_desc} and][]{polletta08c}.}\\
\tablefoottext{b}{$\nu$ is the observed frequency of the CO line fitted
with a single Gaussian. $\nu_1$ and $\nu_2$ are the observed frequencies of
the two CO peaks obtained with a double Gaussian fit, and FWHM, FWHM$_1$,
and FWHM$_2$ are the corresponding full-widths-at-half maximum.}\\
\tablefoottext{c}{Velocity offset between the peaks of the double
Gaussian fit.}\\
\tablefoottext{d}{Intensity derived from each component of the double
Gaussian fit.}\\
\tablefoottext{e}{L$_{CO}$ is the CO line luminosity, in\,\lsun.}\\
\tablefoottext{f}{L$^{\prime}_{CO}$ is the integrated CO luminosity, in
K\,\kms\,pc$^2$.}\\
\tablefoottext{g}{M$_{gas}$ is the gas (H$_2$+He) mass.}\\
\tablefoottext{h}{SFE is the star-formation efficiency defined as
L(FIR)/L$^{\prime}_{CO}$ (see~\S~\ref{sfe}).}
}
\end{table}

\section{Selected sources: obscured QSOs at $z$$\gtrsim$3.4}\label{src_desc}

The targets selected for this study are \object{SWIRE2
J022550.32$-$042149.6} (SW022550 hereafter) at $z$=3.867$\pm$0.009, and
\object{SWIRE2 J022513.90$-$043419.9} (SW022513 hereafter) at
$z$=3.427$\pm$0.002. Both sources are situated in a region common to the
SWIRE~\citep{lonsdale03}, XMDS~\citep{chiappetti05}, CFHTLS
Deep\footnote{http://www.cfht.hawaii.edu/Science/CFHTLS/}, and
UKIDSS~\citep{dye06,lawrence07} surveys. They benefit from a wealth of
high-quality multi-$\lambda$ data, from X-ray to radio wavelengths, and
optical and near-infrared (NIR) spectroscopic data, including VLT/SINFONI
imaging spectroscopy~\citep[see][for a description of their multi-wavelength
data and properties]{polletta08c,nesvadba11b}. A fit to their
multi-wavelength spectral energy distribution (SED) indicates a dominant AGN
component in the MIR and a starburst at longer wavelengths. Their optical
and near-infrared spectra display strong narrow high ionization emission
lines typical of obscured AGNs, and fainter blueshifted broad
components~\citep[see ][]{polletta08c,nesvadba11b}. They are very luminous
QSOs (L$_{bol}\sim$10$^{47}$\,\ergs) with high MIR/optical flux ratios ($\nu
F_{24\,\mu m}/\nu F_{z}$$>$30), and faint and hard X-ray emission typical of
heavily absorbed sources. The estimated stellar masses for both systems are
a few times 10$^{11}$\,\msun, and the star-formation rates (SFRs)
$\sim$500--3000\,\msun/yr~\citep{polletta08c}. In Table~\ref{co_params} we
report the total IR luminosity [L(IR)], estimated SFR, stellar masses and
redshifts for both targets. The IR luminosity [L(IR)] is the 8--1000\,$\mu$m
luminosity in \lsun\ derived after removing the AGN contribution. The
star-formation rate, SFR, is derived from L(IR), using the following
prescription: SFR(\msun/yr) =
L(IR)/(5.8$\times$10$^9$\lsun)~\citep{kennicutt98a}. The stellar mass,
M$_{stellar}$, is derived from the H-band (1.6\,\um\ in the rest-frame)
luminosity assuming the L(H)/M$_{stellar}$ ratio reported
by~\citet{seymour07}.  Note that L(IR), and thus the SFR, are not well
constrained due to the lack of data between 24\,\um, and 1.2\,mm. For more
details on these estimates see~\citet{polletta08c}. The bright
($\sim$5\,mJy, see Table~\ref{co_params}) millimeter fluxes and accurate
redshifts of SW022550 and SW022513 motivated CO observations with the
Plateau de Bure Interferometer (PdBI) of both targets. The observations are
described in the next section, the CO line properties in~\S~\ref{CO_data},
and the molecular gas dynamics and implications in \S~\ref{discussion}.

\section{PdBI observations and data reduction}\label{observations}

Observations with the IRAM PdBI were carried out on August 3--9, 2009, in the D
configuration. The 3-mm receivers were tuned to the central observed
frequency according to the spectroscopic redshifts. We centered the 1\,GHz
(about 3000\,km\,s$^{-1}$) observation band at the redshift derived from the
UV rest-frame lines for SW022550, $z$=3.867$\pm$0.009\footnote{The redshift
of SW022550 has been revised after the PdBI observations were carried out.
The UV spectrum presents two main sets of emission lines, one at $z$=3.867,
and another, dominated by the stronger narrow emission lines, at
$z$=3.876~\citep[see][]{polletta08c}.}, and at the redshift derived from the
optical rest-frame lines for SW022513, $z$=3.427$\pm$0.002. At these
redshifts, the $J$=4-3 line is the only CO line ($\nu_{rest}$=461.0408\,GHz)
within the 3\,mm atmospheric window. The $J$=4-3 CO line was expected at a
frequency of 94.6113\,GHz, and 104.261\,GHz, in SW022550, and SW022513,
respectively.

Both sources were observed under conditions of reasonable atmospheric
phase stability (seeing$\simeq$1.2\arcsec--1.5\arcsec) and transparency
(pwv$\simeq$3--10\,mm) with one track per source and using all six antennas
for a total integration time of 5.3\,hrs, and 8.9\,hrs, for SW022550, and
SW022513, respectively. The data reduction was carried out in two stages
using the IRAM GILDAS
software\footnote{http://www.iram.fr/IRAMFR/GILDAS.}~\citep{guilloteau00}. 
The raw data were calibrated using the {\sc clic} package developed at IRAM.
First, calibration anomalies owing to meteorological conditions or electronics
were rejected. After calibration, the visibility data, i.e., $uv$ tables,
were generated using {\sc clic}. Based on these $uv$ tables, CO maps and
spectra were extracted using the IRAM MAPPING software version 4 .

\begin{figure*}
\centering
\includegraphics[keepaspectratio='true',width=9.cm,angle=0]{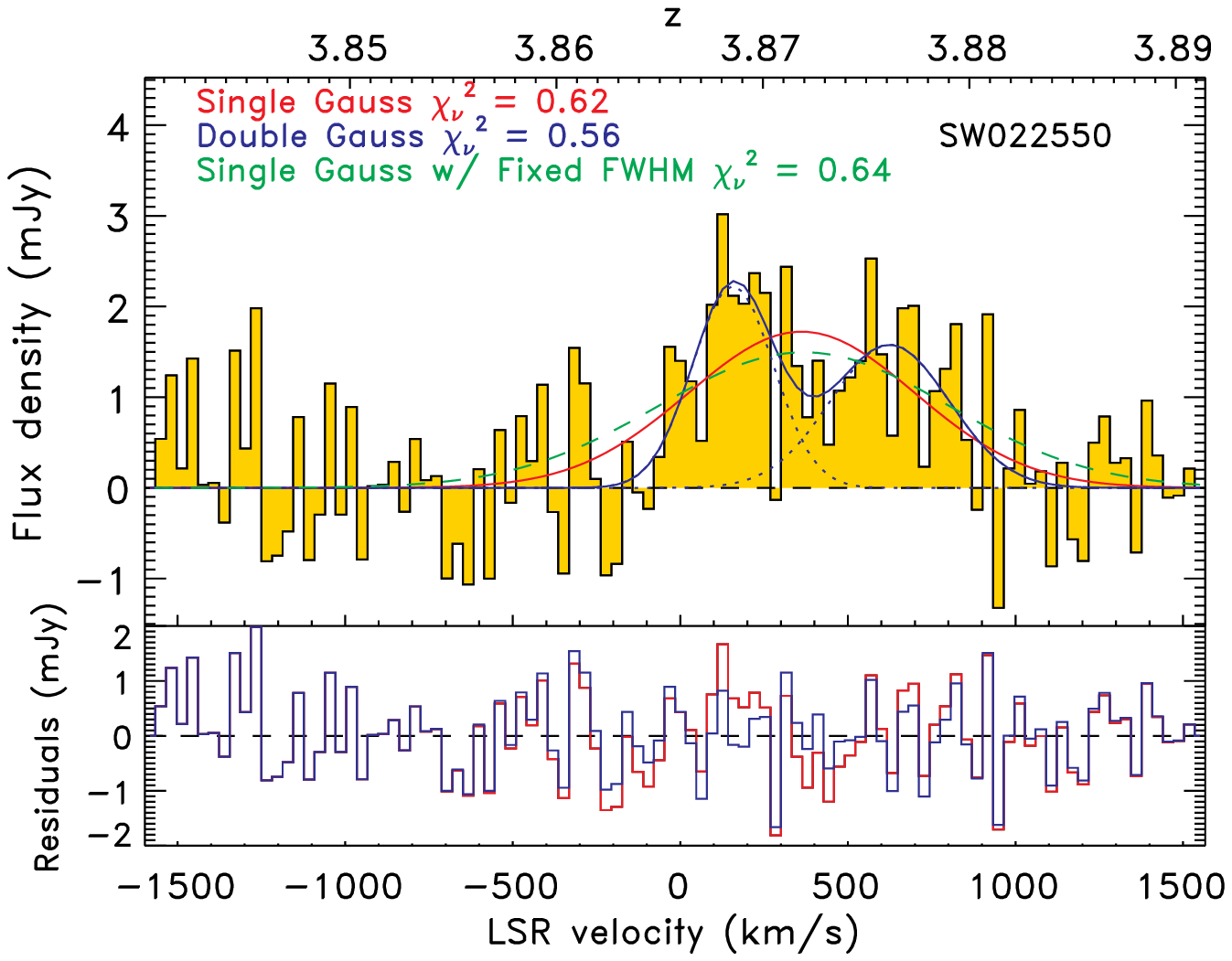}
\includegraphics[keepaspectratio='true',width=9.cm,angle=0]{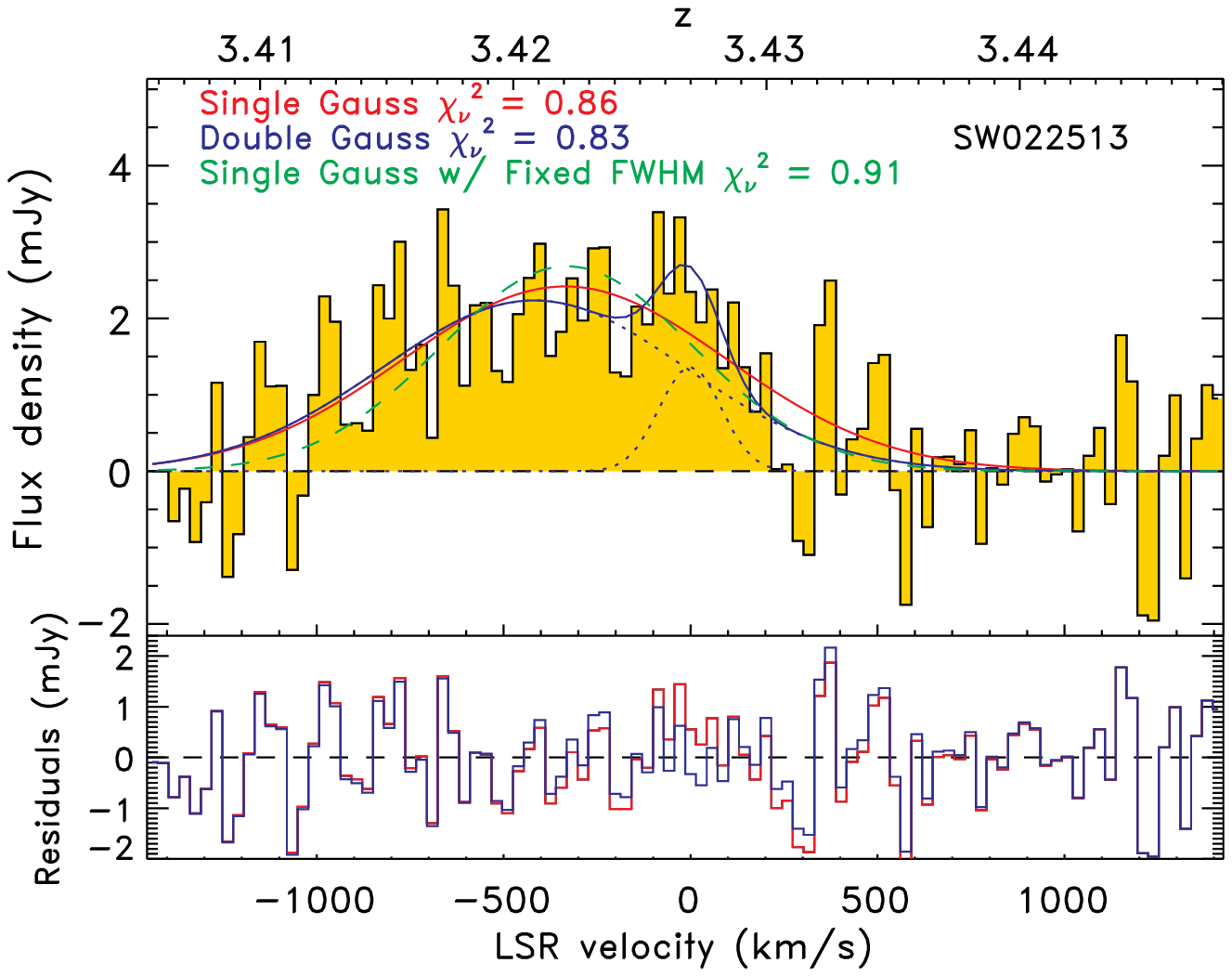}
\caption{{\small CO(4--3) spectrum of the CO emission in the brightest pixel
of the central source in SW022550 ({\it left panel}) and SW022513 ({\it
right panel}). The LSR velocity scale is relative to $z$=3.866 for SW022550,
and $z$=3.427 for SW022513. The lines represent the best-fits obtained with
a single Gaussian (red solid line), and with a double Gaussian (each
component is show with a dotted blue line and the sum with a solid blue
line). The green dashed line represents a single Gaussian fit with FWHM
equal to the best-fit value obtained for the other source, i.e.,
FWHM\,=\,800\,\kms\ in SW022513 and FWHM\,=\,1020\,\kms in SW022550. The
best-fit parameters are listed in Table~\ref{co_params} and the
reduced-$\chi^2$ are annotated on the upper left corner. The residuals
obtained from the single and the double Gaussian component fits are shown in
the bottom panels with a solid red and blue line, respectively.}}
\label{co_fits}
\end{figure*}

The synthesized, clean beam is elongated, roughly 8.4\arcsec
$\times$4.8\arcsec\ for SW022550, and 6.5\arcsec $\times$4.3\arcsec\ for
SW022513. Before producing the integrated CO emission-line maps, the data
were rebinned by a factor of 4, i.e., from the initial channel width of
2.5\,MHz to a width of 10 MHz. Our final, spectrally smoothed data cube for
SW022550 has rms\,=\,0.1\,mJy\,beam$^{-1}$ for a channel width of 32\,\kms. 
For SW022513, we reached rms\,=\,0.17\,mJy\,beam$^{-1}$ for a channel width
of 29\,\kms. Spectra extracted from the brightest pixel of each source are
shown in Figure~\ref{co_fits}.

\begin{figure}
\centering
\includegraphics[keepaspectratio='true',width=7.cm,angle=0]{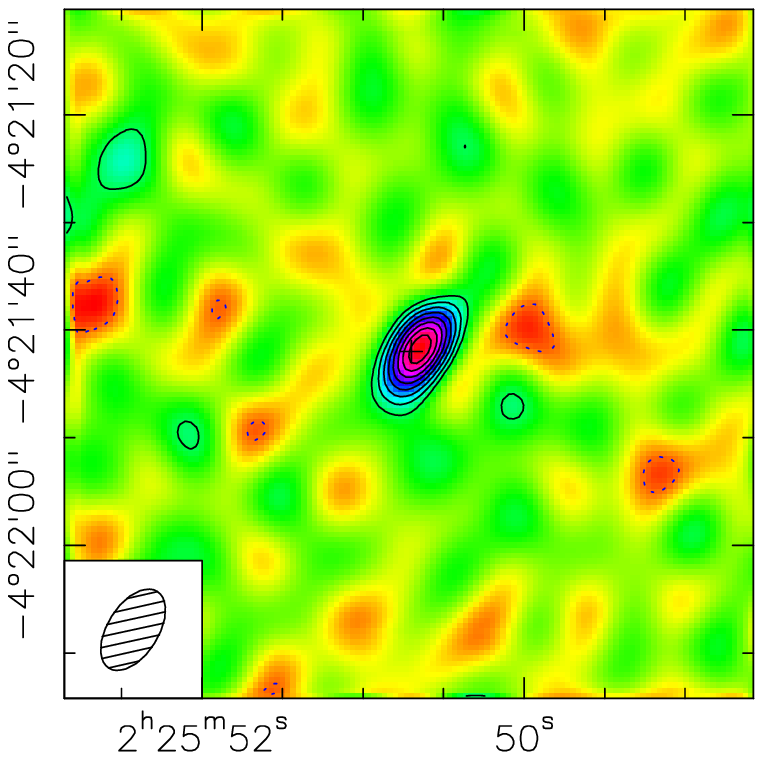}
\includegraphics[keepaspectratio='true',width=7.cm,angle=0]{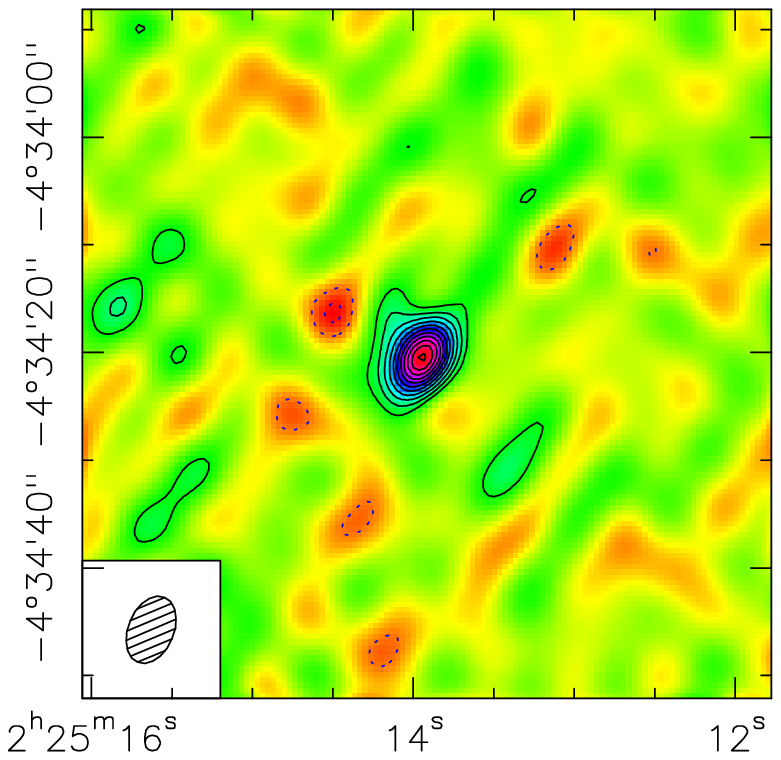}
\caption{{\small CO(4--3) velocity-integrated map of SW022550 ({\it top
panel}) and of SW022513 ({\it bottom panel}). CO contours are plotted with a
1\,$\sigma$ step and start at 2\,$\sigma$ (1\,$\sigma$=0.13, and
0.19\,Jy\,\kms\,beam$^{-1}$ for SW022550, and SW022513, respectively). The
synthesized PdBI beams are shown in the inset as hatched ellipses.}}
\label{co_maps}
\end{figure}

We also constructed CO emission line maps from these cubes by integrating
over a velocity width ranging from $-$20\,\kms\ to 870\,\kms\ for SW022550,
and from $-$980\,\kms\ to 200\,\kms\ for SW022513, both centered on the peak
CO line emission. These maps are shown in Figure~\ref{co_maps}, where
contours are plotted with 1$\sigma$ step size and the first CO contour
starts at 2$\sigma$, and $\sigma$=0.13\,Jy\,\kms\,beam$^{-1}$ in SW022550,
and $\sigma$=0.19\,Jy\,\kms\,beam$^{-1}$ in SW022513.

\section{Molecular gas measurements}\label{CO_data}

The CO-related measurements for our two selected sources
are reported in Table~\ref{co_params} and described here.

\subsection{CO line intensity}\label{co_intensity}

In both SW022550 and SW022513, strong CO lines are detected with integrated
fluxes of I$_{CO}$\,=\,1.5$\pm$0.3\,Jy\,\kms, and 2.6$\pm$0.4\,Jy\,\kms,
respectively. No continuum is detected in either source. The estimated
3$\sigma$ upper limits in the continuum maps of SW022550 and SW022513 are
0.3mJy/beam and 0.5mJy/beam, respectively. Both sources appear compact in
our observations with beam size $>$4\,\arcsec\ corresponding to
super-galactic scales, i.e., $>$30\,kpc at $z$=3.4. The estimated CO(4--3)
luminosity, L$_{CO}$, is derived from the line intensity I$_{CO}$ and the
following equation~\citep{solomon97}:
L$_{CO}$\,=\,1.04$\times$10$^{-3}\times$I$_{CO}\times \nu_{obs}\times
D_L^2$, where $\nu_{obs}$ is the line-observed frequency in GHz, and $D_L$
the luminosity distance in Mpc. The integrated CO luminosity,
L$^{\prime}_{CO}$, derived using the following equation:
L$^{\prime}_{CO}$\,=\,3.25$\times$10$^7 \times$I$_{CO}\times
\nu_{obs}^{-2}\times D_L^2\times (1+z)^{-3}$, is
(5.7$\pm$1.0)$\times$10$^{10}$\,\,K\,\kms\,pc$^{-2}$ in SW022550, and
(8.3$\pm$1.2)$\times$10$^{10}$\,\,K\,\kms\,pc$^{-2}$ in
SW022513~\citep[see][and Table~\ref{co_params}]{solomon97}.

Assuming an L$^{\prime}_{CO(1-0)}$/L$^{\prime}_{CO(4-3)}$ ratio equal to
unity, as observed in high-$z$ type I
QSOs~\citep{riechers06,weiss07}\footnote{Note that the
L$^{\prime}_{CO(1-0)}$/L$^{\prime}_{CO(4-3)}$ ratio typically ranges from
0.65 to 1.73 with high-$z$ type I QSOs being characterized by values close
to unity~\citep{riechers06,weiss07}, and submillimeter galaxies by lower
value~\citep[e.g.][]{hainline06,ivison11,harris10,frayer11}.\label{fn_co}},
the total H$_2$+He molecular gas mass can be estimated as
M$_{gas}$\,=\,M(H$_2$+He)\,=\,$\alpha$ L$^{\prime}_{CO(1-0)}$, where
$\alpha$ is the CO luminosity to gas mass conversion factor. We adopt
$\alpha$\,=\,0.8\,\msun (K\,\kms\,pc$^2$)$^{-1}$, which has been estimated
from local ultra-luminous infrared
galaxies~\citep[ULIRGs;][]{solomon97,downes98}, and is commonly adopted for
high-redshift ULIRGs~\citep{solomon05,tacconi08,bothwell10,yan10}. This
value also includes a correction factor of 37\% for helium. With these
assumptions, the resulting total gas masses, including only the measurement
errors and not the astrophysical uncertainties, are
(4.5$\pm$0.8)$\times$10$^{10}$\,\msun\ for SW022550, and
(6.7$\pm$0.9)$\times$10$^{10}$\,\msun\ for SW022513. Note that these
molecular gas mass estimates are likely to be underestimated because the
$J$=4--3 transition might not trace the bulk of the molecular gas, and the
estimates are probably uncertain by a factor of at least 2~\citep[for a
discussion on the hypothesis behind this assumption and possible
uncertainties see][]{ivison11}. Table~\ref{co_params} lists the integrated
CO line fluxes, velocity widths, and derived CO luminosities and gas
(H$_2$+He) masses.

\subsection{CO line profiles}\label{co_profiles}

We investigate the gas kinematics of our sources by fitting the line profile
with a single and a double Gaussian model, and use reduced-$\chi^2$ to
estimate the goodness-of-fit for each model (see also Fig.~\ref{co_fits}).
For SW022550 a double Gaussian fit yields the best fit, $\chi_{\nu}^2$=0.56
compared to $\chi^2_{\nu}$=0.62 with a single Gaussian. The widths of the
two Gaussian components are consistent with each other and, indeed, a double
Gaussian model with equal line widths yields lines with full-width-at-half
maximum of 340$\pm$60\,\kms\ and an equally acceptable fit, with the same
$\chi_{\nu}^2$. For SW022513, fits with one and two components give an
equally acceptable fit, $\chi^2_{\nu}$$\sim$0.85, however, the best
two-component fit in this case implies very different line widths,
FWHM\,=\,200$\pm$150\,\kms\ and FWHM\,=\,950$\pm$150\,\kms\ for the narrow
and broad component, respectively. The broad component is blueshifted by
$-$420\,\kms\ relative to the narrow component. \citet{nesvadba11b} find
that the line can be equally well fit with a narrow and a broad component
that have the same redshift and width as those found for the narrow and
broad components of \hbeta, respectively. All fit results are listed in
Table~\ref{co_params} and shown in Fig.~\ref{co_fits}. We rounded all
results to multiples of 10 km s$^{-1}$.

We emphasize that the profiles of SW022513 and SW022550 are significantly
different. When fitting SW022513 with a single Gaussian as broad as in
SW022550 (and vice versa), the $\chi^2$ worsens (see annotations in
Fig.~\ref{co_fits}). We tried the same exercise for the two-component fit
with fixed widths, but this fit did not converge for SW022550, and produced
a very bad result for SW022513 (not shown in Fig.~\ref{co_fits}). In
addition, a Kolmogorov-Smirnov (KS) test rejects the hypothesis that both
spectra are similar at 98\% confidence. Before applying this test, we
shifted the spectra to the central wavelength of the single Gaussian fit
(corresponding to shifts of $-$364\,\kms\ and +336\,\kms\ for SW022550 and
SW022513, respectively), selected the velocity range 
covered by the lines in both galaxies ($-$1000\,\kms\ to +1190\,\kms) and
resampled the spectra to have the same spectral binning. All fits and
associated $\chi^2_{\nu}$ are reported in Fig.~\ref{co_fits}.

Based on these tests, we conclude that the line profiles of SW022550
and SW022513 are significantly different, suggesting that the molecular gas
in these galaxies could be in different dynamical states, in spite of
their similar SEDs and MIR/FIR luminosities.

\subsubsection{Are the line profiles in our sources typical?}\label{comparison}

Double-horned profiles, as in SW022550, are often found in merging
galaxies, for example submillimeter galaxies~\citep[e.g.
][]{frayer98,neri03,tacconi06,bothwell10}, and may either signalize a
rotating disk or a merger of two reservoirs of molecular
gas~\citep[e.g.,][]{neri03,taniguchi98,sakamoto99,narayanan06,downes98,evans99}.
However, double-peaked profiles with FWHM$\geq$300\,\kms\ per peak are rarely observed.

Lines with FWHM$\sim$1000\,\kms, similar to SW022513 are also rare, and are
often double-peaked \citep[like in 4C41.17 at $z$=3.8;][and in
SMMJ02399-0136;~\citealt{genzel03}]{debreuck05}. \citet{papadopoulos00}
report a line with FWHM=1000\,\kms\ in 4C60.07, which is not associated with
the radio galaxy, but with an infalling satellite galaxy \citep[see
also][]{ivison08}. \citet{coppin08b} report a CO line with FWHM=1090\,\kms\
in the BAL QSO RX\,124913.86-05906.2 at $z$=2.247, which appears to have a
profile not very different from ours, but at lower signal-to-noise ratio,
making the width measurement more uncertain. Thus, the CO line widths of our
sources are not unique, but nonetheless uncommon among high-$z$ galaxies
with bright CO line emission.

To examine whether similar CO line profiles are also observed in other
obscured AGNs, and because no systematic CO studies have been carried out
for this class of objects, we have searched the literature for other
CO-detected obscured AGNs at high-$z$ and collected their main CO
properties. The vast majority of CO-detected sources at high redshift are
classified as SMGs, type I QSOs, ULIRGs, or radio galaxies, even if in a few
cases they host an obscured AGN. We collected all CO-detected sources that,
to our knowledge, exhibit properties typical of obscured AGNs, like narrow
high ionization emission lines in their optical spectrum, or absorption in
the X-rays. We also require that the selected sources are ULIRGs
(L(FIR)$>$10$^{12}$\,\lsun) and not radio-loud objects. We found eight
sources that satisfy these criteria. Because it is not always clear whether
a source contains an obscured AGN, our compilation is not complete. Note
that in case of \object{COSMOS\,J100038+020822}, the optical spectrum is typical of a
broad line AGN, but we include this source in our compilation because its
spectral energy distribution and X-ray spectrum indicate that the source
suffers heavy obscuration.

The list of the eight CO-detected obscured AGNs found in the literature and
their main CO properties (i.e., line width, offset from the systemic
redshift) are reported in Table~\ref{qso2_tab}. In
Table~\ref{qso2_tab} we also report their mean values (without including
our selected sources), and, for reasons of practicality, we recall the same
quantities for our two selected targets.

\begin{table*}[ht!]
\caption{\label{qso2_tab}Main properties of CO-detected obscured AGNs from the
literature}
\centering
\begin{tabular}{lccc cccc cc}
\hline\hline
Object                          & $z_{opt}$ &  CO      & $z_{CO}$ & Offset & FWHM(CO)     &  M$_{gas}$      &  SFR     & SFE\tablefootmark{a} & Ref.\tablefootmark{b} \\
                                &           &  trans.  &          & \kms\  & \kms         &  10$^{10}$\msun & \msun/yr &   &   \\
\hline
SW022550                        &  3.876    &  (4--3)  & 3.8719   & $-$460, 10 & 290$\pm$90, 420$\pm$180 & 4.53           &500--3000 & 35--194 & \nodata \\
SW022513                        &  3.427    &  (4--3)  & 3.4220   & $-$340 &  1020$\pm$110        & 6.66            &500--3000 & 24--108 & \nodata \\
\hline
\object{MIPS\,8342}             &  1.562    &  (2--1)  & 1.5619   &  $-$12  &134$\pm$40, 144$\pm$40\tablefootmark{c}&1.79& 610 &    60   & (1) \\
J100038                         &  1.8288   &  (4--3)  & 1.8269   & $-$200 &   427$\pm$73 &  3.6--5.4       &  1700    &   325   & (2) \\
\object{MIPS\,15949}            &  2.122    &  (3--2)  & 2.1194   & $-$250 &   500$\pm$117&  2.18           &  1370    &    97   & (1) \\
IRAS\,F10214                    &  2.2856   &  (4--3)  & 2.2854   &  $-$18 &   245$\pm$28 &  0.6            &   540    &   719   & (3) \\
\object{MIPS\,16059}            &  2.326    &  (3--2)  & 2.3256   &  $-$36 &   471$\pm$54 &  1.41           &  1280    &   211   & (1) \\
\object{MIPS\,8327}             &  2.441    &  (3--2)  & 2.4421   &     96 &   253$\pm$50 &  1.02           &  1170    &   231   & (1) \\
\object{AMS\,12}                &  2.767    &  (3--2)  & 2.7668   &  $-$16 &   275        &  1.9            &  4470    &  1047   & (4) \\
SMM\,J02399                     & 2.803 &  (3--2)  & 2.8076   & $-$520, +230 & 220, 220\tablefootmark{d}&6.0 & 500     &    90   & (5) \\ 
\hline
Mean\tablefootmark{e}           &2.27$\pm$0.32&\nodata &2.27$\pm$0.32&$-$73$\pm$94&450$\pm$180& 2.4$\pm$1.4  &1455$\pm$815 & 347$\pm$268 & \nodata \\
\hline
\end{tabular}
\tablefoot{J100038 stands for COSMOS\,J100038+020822, IRAS\,F10214 for
\object{IRAS F10214+4724}, and SMM\,J02399 for the L1 component of
\object{SMM\,J02399$-$0136}. All
sources, but COSMOS\,J100038, are classified as type 2 QSOs based on their
optical spectrum.\\
\tablefoottext{a}{SFE is the star-formation efficiency, in units of \lsun/(K\,km\,s$^{-1}$\,pc$^2$).}\\
\tablefoottext{b}{(1):~\citet{sajina08,yan10}, (2):~\citet{aravena08},
(3)~\citet{solomon05}, (4)~\citet{martinez09},
(5)~\citet{genzel03,iono09,frayer98}.}\\
\tablefoottext{c}{FWHM of each component of a double-peaked CO line. A single Gaussian
fit yields FWHM\,=\,325\,\kms.}\\
\tablefoottext{d}{The velocity offsets and FWHMs refer to the two peaks of a
double-horned CO line. The velocity offsets are given with respect to
$z_{CO}$=2.808~\citep{frayer98}. The full line covers $\sim$1100\,\kms\ in
wavelength.}\\
\tablefoottext{e}{Mean values and average deviation from the mean
obtained from the eight sources from the literature.}
}
\end{table*}

The obscured AGN redshifts range from 1.6 to 2.8 and the CO lines refer
to the $J$\,=\,(2--1), (3--2), or (4--3) transitions. In case multiple transitions
were available, we considered the one closest to $J$=(4--3). Two out of
these eight sources show a double-horned CO profile. The line widths,
based on single Gaussian fits, range from 245 to 1100\,\kms, similar to
the range observed in SMGs and type 1 QSOs, i.e., 200--1000\,\kms.

Previous studies of the line profiles of CO bright sources at high-$z$,
i.e., SMGs and type I QSOs, indicate that their CO lines are characterized by
similar widths~\citep[$<$FWHM$>$=550$\pm$180\,\kms\ in QSOs, and
$<$FWHM$>$=530$\pm$110\,\kms\ in SMGs;][]{coppin08b,wang10}, but 
double-peaked profiles are more common among SMGs than among type I QSOs.
This difference is attributed to a closer alignment of the gas plane with
the line-of-sight in SMGs than in QSOs. The mean CO line width in the
selected sample of eight obscured AGNs is 450$\pm$180\,\kms\ (540$\pm$260\,\kms\
including our two sources), consistent with the mean CO line width observed
in SMGs and type 1 QSOs~\citep{coppin08b}. Thus, the CO line widths are all
consistent among these three main classes of high-$z$ galaxies, but in terms
of line profile, our sources are more similar to SMGs than to type I QSOs.

Velocity offsets from the systemic redshift, assumed to be the optical
redshift, are common and range from about $-$500\,\kms\ to +100\,\kms.
Blueshifted CO lines are observed in the vast majority of sources, and in
three out of eight sources the velocity offset reaches a few hundreds of
\kms\ like in our sources.

Although the widths of the CO lines in SW022550 and in SW022513 are not
unique among the sources from the literature, they are among the largest
ever observed. Interestingly, the largest CO lines are only observed in
sources that also contain a powerful AGN, suggesting that the AGN might play
a major role in broadening the CO line. The hypothesis that the AGN might be
responsible for the CO line width, at least in SW022513, will be discussed
in more detail in \S~\ref{wind}.

\subsection{Are the sources spatially resolved?}\label{pair_case}

\begin{figure}
\centering
\includegraphics[keepaspectratio='true',width=9cm,angle=-90]{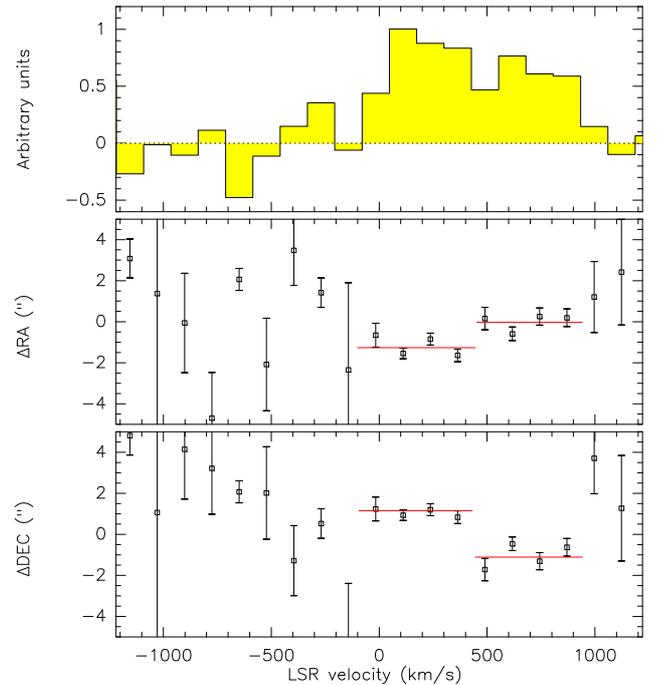}
\caption{ {\it Top panel:} SW022550 CO(4--3) line profile resampled to a
frequency resolution of 40\,MHz ($\sim$130\,\kms). Right
ascension ({\it middle panel})  and declination ({\it bottom panel}) offsets in
arcsec relative to the two peaks seen in the CO(4--3) line profile.}
\label{spatial_offset}
\end{figure}

\begin{figure}
\centering
\includegraphics[keepaspectratio='true',width=9cm,angle=0]{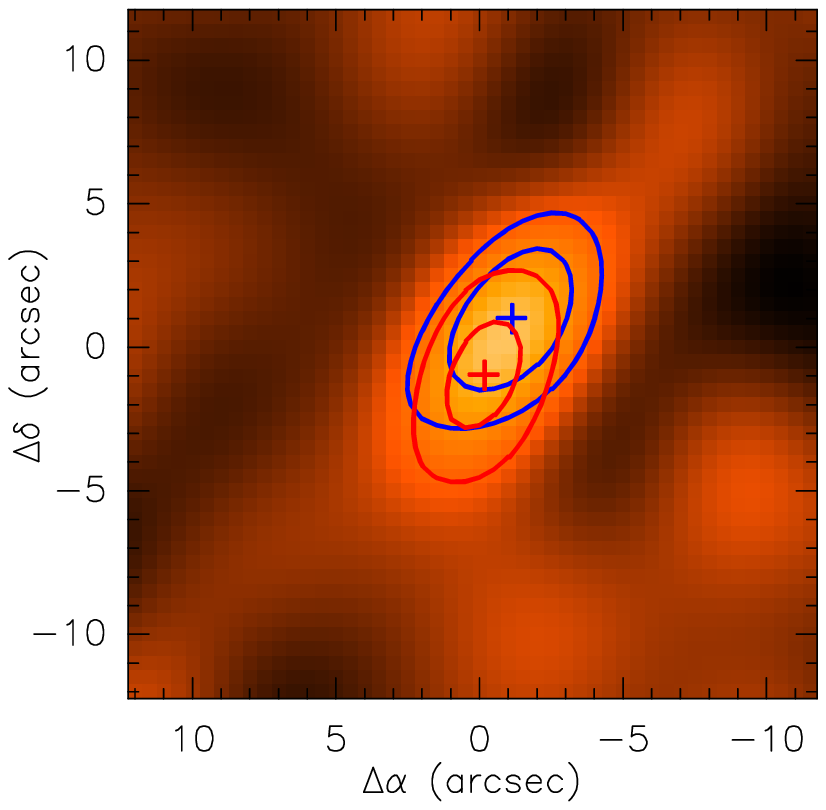}
\caption{ {\small CO(4--3) velocity-integrated map of SW022550 and CO
contours relative to the two peaks. The blue and red contours and crosses
correspond to the blue and red components, respectively. Contours are shown
at 3 and 5$\sigma$ with $\sigma$=0.11 Jy\,\kms\,beam$^{-1}$. The maps
relative to each peak have been obtained by integrating two adjacent
160\,MHz ($\sim$510\,\kms) wide bins in correspondence of the velocity
intervals marked by the red horizontal lines in Fig.~\ref{spatial_offset}.
Crosses mark the peak of emission and their size corresponds to 1\arcsec.}}
\label{red_blu_map}
\end{figure}

Our data have beam sizes of $\gtrsim$4\arcsec\ (corresponding to
$\simeq$30\,kpc at $z$=3.4), larger than typical galactic scales.
Nonetheless, in SW022550, we may have found a small spatial offset between
the red and the blue Gaussian component of about 1\arcsec\ and 2\arcsec\ in
right-ascension and declination, respectively (Fig.~\ref{spatial_offset}).
We estimated this offset from channel maps extracted from the blue and red
peak of the line extracted from a cube with a channel width of 40\,MHz
($\sim$130\,\kms). The peak emission and relative positions of the red
and blue components are shown in the integrated map in
Figure~\ref{red_blu_map}.

With our beam size, 8.4\arcsec$\times$4.8\arcsec, and SNR$\sim$7 for each
peak (corresponding to a total SNR$\sim$10), the relative positional
accuracy of our data is 0.6\arcsec$\times$0.3\arcsec\ in right ascension and
declination, respectively\footnote{The 1$\sigma$ positional accuracy is the
size of the synthesized beam in right ascension or declination divided by
twice the SNR, i.e., 1$\sigma$ in RA is
8.4\arcsec/(2$\times$7)=0.60\,\arcsec, and 1$\sigma$ in Dec is
4.8\arcsec/(2$\times$7)=0.34\,\arcsec.}. Thus, the offset is marginally
significant at a 2$\sigma$-level along right ascension, and is significant
at 6$\sigma$-level along declination. This suggests that the molecular gas
in SW022550 may extend over 2.2$\pm$0.7\arcsec\ (16$\pm$5\,kpc). Since we
cannot distinguish if the gas is in two spatially separate or in one single,
extended component, we cannot differentiate between the disk and the (two
components) merger hypothesis. This would require observations at higher
spatial resolution. Extended molecular disks with sizes of $\sim$10 kpc have
been already observed, e.g., in the low-redshift galaxy
3C\,293~\citep{evans99} and in the $z$$=$1.5 SMG
Lockman\,38~\citep{bothwell10}. Interestingly, extended molecular disks at
low redshift are often associated with (advanced) mergers~\citep{downes98,
evans99}, but several cases of extended disks in non-mergers are also
known, especially at high-redshift~\citep{tacconi10,daddi10b}.

We currently have no evidence for multiple components in our multi-band
imaging at shorter wavelengths, however, this does not necessarily rule out
a merger of two separate galaxies. Our highest-resolution data ($<$1\arcsec)
are at wavelengths $\le$1\,\micron\ where obscuration might be important,
and where the galaxy could be outshone by the AGN. Optical and
near-infrared continuum observations at high-$z$ can also be more difficult
than line detections because the cosmological surface-brightness dimming is
a stronger function of redshift for continuum than for line emission
($\propto(1+z)^4$ compared to $(1+z)^5$). It is also possible that the gas
is intrinsically more extended than the continuum.

Fitting a circular Gaussian to the CO(4-3) velocity integrated
visibilities of SW022513 gives a 3$\sigma$ upper limit on the size of
2.4\arcsec\ (18\,kpc at $z$=3.4). The continuum morphology at optical and
near-infrared wavelengths is compact, but SW022513 has a narrow-line
region seen with VLT/SINFONI that extends over
2.5\arcsec$\times$1.6\arcsec\ \citep{nesvadba11b}, consistent with our upper
limit. SW022513 is also resolved at 1.4 GHz with a size of
4.1\arcsec$\times$1.8\arcsec\ \citep{bondi03}, larger than our upper limit.

\subsection{Dynamical masses}\label{dyn_masses}

A common assumption made in studies of the CO line emission of high-redshift
galaxies is that the molecular gas is virialized and rotating in the
large-scale gravitational field of the galaxy, in particular when the line
profile is double-horned. Recent observations of the submillimeter galaxy
SMM\,J2131-0102~\citep{danielson10}, suggest however, that the CO kinematics
of dusty starburst at high redshift can be significantly more complex than
that. Owing to the strong magnification by a gravitational lens, it has been
possible to observe three kinematically separate components in this galaxy,
with brightness in each component depending on the observed transition. This
may suggest that the CO kinematics of dusty starbursts at high redshift are
significantly more complex than they may appear at the generally low
signal-to-noise ratios that can be achieved for galaxies that are not
strongly lensed. This suspicion may be particularly appropriate for our
targets, where observations at different (and observationally less
challenging) wavebands suggest a great level of kinematic
complexity~\citep{nesvadba11b}, and may hint at a non-gravitational origin
of the CO kinematics, at least in the case of SW022513. We will discuss
these alternative interpretations in \S~\ref{discussion}. Irrespective of
these concerns, here we adopt the more traditional interpretation that the
CO line profiles are dominated by gravity, and that the gas is virialized.
Since dynamical mass estimates depend on the system geometry, we assume that
the gas is virialized~\citep[but see][]{ivison08} and consider two cases,
the gas distributed in a rotating disk, or associated with two merging
galaxies. The estimated dynamical masses, M$_{dyn}^{disk}$ and
M$_{dyn}^{merger}$, are reported in Table~\ref{co_params}.

For a rotating disc, the dynamical mass M$_{dyn}^{disk}$ is derived
following~\citet{neri03}, as M$_{dyn}^{disk}\ sin^2(i)$\,=\,2.33$\times$
10$^5\times\Delta$V$^2\times R$ [\msun], where $i$ is the inclination of the
putative disk with respect to the plane of the sky, $\Delta$V is the
difference in velocity between the two peaks in the CO profile, or, in case
of a single peak, the observed FWHM in \kms divided by 2.4, and $R$ is the
disk radius in kpc. We use the velocity offset between the two peaks in
SW022550, $\Delta$V\,=\,460\,\kms\, and
FWHM/2.4\,$\simeq$\,1020/2.4\,$\simeq$\,425$\pm$50\,\kms\ in SW022513.  For
SW022550 we adopt as radius $R$ half of the positional offset between the
two CO line peaks, i.e., $R$=8$\pm$2.5\,kpc. Because the CO is unresolved
in our maps, we consider a range of fiducial values for $R$ for SW022513,
corresponding to the lowest and highest $R$ values measured in similar
sources at high-$z$. Disk sizes typically range from 0.5 to 5\,kpc in
QSOs~\citep{carilli03,walter04,weiss07,aravena08,riechers08,riechers09a} and
from 2 to 8\,kpc in submillimeter
galaxies~\citep{neri03,greve05,tacconi06,tacconi08,swinbank10,carilli10,ivison11}.
Thus, we consider a range of values between 0.5 and 8\,kpc for the radius
$R$ in SW022513. With these assumptions, the estimated dynamical mass in a
disk, M$_{dyn}^{disk}$, is (4$\pm$2)$\times$10$^{11}$\,sin$^{-2}(i)$\,\msun\
in SW022550, and ranges from (2.0$\pm$0.5)$\times$
10$^{10}$\,sin$^{-2}(i)$\,\msun\ to (34$\pm$8)$\times$
10$^{10}$\,sin$^{-2}(i)$\,\msun\ in SW022513.

An alternative explanation is a merger with highly disturbed gas kinematics
driven by two gravitational potentials, each associated with a merging
galaxy. In this scenario, and again following~\citet{neri03}, the dynamical
mass can be derived as
M$_{dyn}^{merger}$$sin^2(i)$=4.2$\times$10$^4\times$FWHM$^2$$\times$$R$
[\msun], where $i$ is the inclination of the plane defined by the two
merging galaxies with respect to the plane of the sky, FWHM the
full-width-at-half maximum of the CO line in \kms, and $R$ half of the
projected distance between the two galaxies in kpc. We adopt
FWHM=800$\pm$120\,\kms\ for SW022550, and 1020$\pm$110\,\kms\ for SW022513,
and assume that the distance $R$ is 8$\pm$2.5\,kpc in SW022550, and
(2--8)\,kpc in SW022513. In this case, we find dynamical masses that are
consistent with those obtained in case of a disk, i.e.,
M$_{dyn}^{merger}$=(2$\pm$1)$\times$ 10$^{11}$\,sin$^{-2}(i)$\,\msun\ for
SW022550, and a range of M$_{dyn}^{merger}$ going from (9$\pm$2)$\times$
10$^{10}$\,sin$^{-2}(i)$\,\msun\ to (35$\pm$8)$\times$
10$^{10}$\,sin$^{-2}(i)$\,\msun\ for SW022513.

In both scenarios, the largest disk radii or merger separation yield
dynamical masses that correspond to the upper end of the mass function at
high redshift~\citep{seymour07}. Furthermore, they would be even higher if
the systems were less inclined ($i\leq$90\deg). This implies that our
sources might be among the most massive galaxies at their epoch, and low
inclinations are disfavored, unless the molecular gas is less extended than
assumed.

\section{Discussion}\label{discussion}

In this section, we analyze the CO emission line luminosities and profiles
to determine what governs the gas dynamics. We will also compare the
molecular gas properties of our targets with those of other CO-detected
sources at high redshift. This comparison will be discussed in light of the
postulated evolutionary scenario that links SMGs and
QSOs~\citep[e.g.][]{sanders88,granato06,dimatteo05}.

\subsection{Gas dynamics}\label{gas_dyn}

The most plausible scenarios that might explain the observed CO kinematics
(see \S~\ref{co_intensity}) are a rotating disk, a galaxy merger, or an
outflow. Interpreting the gas dynamics of high-$z$ galaxies is very
challenging if we exclusively rely on CO observations. We therefore include
additional constraints from our multi-wavelength data sets to investigate
which hypothesis is overall astrophysically most plausible.

We argued in \S~\ref{co_profiles} that the CO line profiles of SW022550 and
SW022513 are significantly different in spite of their similar gas masses
and FIR/MIR luminosities. The CO line profile in SW022550 is characterized
by two relatively symmetric peaks, each being narrower than the velocity
offset of $\Delta$v=460$\pm$90\,\kms\ between them. SW022550 has a classical
double-horned profile, which may either indicate a merger or a rotating
disk; at any rate, gas kinematics driven by gravity. The spatial offset
(16$\pm$5\,kpc, see~\S~\ref{pair_case}) between the red and the blue peak
further suggests that the gas could be in a moderately to highly inclined
disk that extends over galactic scales or beyond. In this case, the circular
velocity, v$_c$$\geq$230\,\kms, would be consistent with those of massive,
extended stellar disks at low redshift like the Milky Way and place SW022550
amongst massive high-redshift galaxies,
M$_{dyn}$=(4$\pm$2)$\times10^{11}$\,\msun, without implying an unusually
large mass.

A highly inclined disk could also explain the obscuration seen at
optical wavelengths. Evidence for obscuration from dust in the host galaxy
(rather than, or in addition, to a circumnuclear torus) has been found in
several AGN, including obscured QSOs at high redshift
~\citep{keel80,lawrence82,rigby06,ogle06,martinez06a,brand07,sajina07b,polletta08a}.
This scenario agrees with the standard idea that in obscured QSOs absorbing
dust and gas intersect the line of sight.

The disk hypothesis is attractive, but not necessarily unique. Apart from
mergers, double-peaked profiles can also be produced by gas plumes produced
in interacting galaxies or clouds and filaments in dense large-scale
structures~\citep[see e.g., the radio galaxies 4C\,60.07 and TXS0828+193;
][]{papadopoulos00,ivison08,nesvadba09a,tacconi08}. However, these 
configurations appear to be rare and have only been identified in
individual, unusual cases.

Our VLT/SINFONI imaging spectroscopy shows that gravity is not the sole
driver of the overall ISM kinematics in SW022550. The \oiiia\ profile is
much broader than the CO profile, FWHM=2200$\pm$180\,\kms, with a single
identifiable peak. Such broad line profiles are commonly interpreted as gas
stirred up and accelerated by the AGN in the inner narrow-line region
\citep[e.g.,][]{komossa08}. This line probes warm ionized gas that is highly
excited, but low in mass, unlike the CO lines. Thus, since the
molecular gas is apparently less perturbed than the ionized gas, the AGN
most likely does not affect the bulk of the ISM in SW022550. This conclusion
holds irrespective of whether the molecular gas is in a disk or driven by a
merger.

The situation is different in SW022513. The CO line profile is very broad
and can either be fitted with a single or a double set of Gaussian profiles.
However, unlike the double-peaked profile of SW022550, the two-component fit
to SW022513 is very asymmetric, with a narrow (FWHM=200$\pm$150\,\kms) line
to the red ($z_{CO,red}$=3.4270$\pm$0.0007), and a much broader
(FWHM=950$\pm$150\,\kms) component blueshifted by $-$420$\pm$100\,\kms\ 
($z_{CO,blue}$=3.4208$\pm$0.0013) relative to the narrow one. The width of
the broad component is consistent with that obtained by a single-component
fit (FWHM=1020$\pm$110\,\kms). In addition, the CO line profile is well
matched by the profile of the \hbeta\ line in the VLT/SINFONI data
cube~\citep[the line includes a narrow component at
$z_{H\beta}$=3.4247$\pm$0.001 with FWHM(\hbeta)=370$\pm$40\,\kms, and a
broad one at $z_{H\beta}$=3.422$\pm$0.001 with
FWHM(\hbeta)=1000$\pm$70\,\kms;][ see their Figure~7]{nesvadba11b} with the
sole exception that the line core of the narrow \hbeta\ component is more
prominent relative to the broad \hbeta\ component compared to the two CO
components. Line widths and redshifts are consistent for CO and \hbeta\ within
the uncertainties.

Unlike SW022550, SW022513 does not have a clear signature of a large-scale
velocity gradient. It is unresolved in the CO map with a 3$\sigma$ size
limit of 2.4\arcsec, suggesting that the molecular gas distribution could be
more compact or as extended as the ionized gas resolved with
VLT/SINFONI. We also do not find a symmetric double-peaked profile as for
SW022550. This suggests that the molecular gas in SW022513, if it is in a
rotating disk, would most likely be seen nearly face-on. In this case, the
line width would not be dominated by the large-scale velocity gradient, but
by motion along the disk normal, i.e., turbulent motion within the disk.
Likewise, to explain this profile with a blend of two lines from a merging
galaxy pair, we would need to postulate a very specific geometry, where the
two galaxies are nearly aligned with the line of sight, and have very
similar line cores, otherwise the profile would be double-peaked. Both
hypotheses require significant fine-tuning, and are therefore, although
fundamentally possible, not very satisfactory. We are not aware of any
example in the literature where a similar line profile was found in either a
merger or a disk.

An alternative interpretation, motivated by the similarity between the
kinematics of molecular and ionized gas and recent discoveries of
molecular outflows in several nearby AGN and starburst galaxies
\citep[][]{sakamoto06,iono09,fischer10,feruglio10,chung11,
alatalo11} is that the broad molecular component could be from gas
that is being stirred up and accelerated by the AGN \citep[see
also][]{nesvadba11b}. This is also supported by N-body/SPH models of
massive galaxy mergers, which demonstrate that line widths with
FWHM$\sim$1000\,\kms\ can only be produced through AGN
winds~\citep{narayanan08a}. We will further investigate this hypothesis in
the next section.

\subsection{A molecular wind as a signature of AGN feedback?}\label{wind}

It has been known for more than two decades that AGN and intensely
star-forming galaxies can drive outflows of warm ionized and neutral gas
\citep[e.g.,][]{vanbreugel84,heckman90}, but it has only
recently been recognized that these winds can also have a molecular
component~\citep[e.g.,][]{iono07}. Consequently, much of what we know about
these winds comes from observations of optical line emission of warm ionized
gas and radio observations of neutral hydrogen. A prime signature of these
winds are broad, blueshifted wings in the integrated spectra of HI
absorption lines and emission lines of ionized
gas~\citep[e.g.,][]{crenshaw99,kraemer05,morganti03,nesvadba06,alexander10}.
Comparison with stellar absorption lines for many types of AGN show that the
narrow components of these lines are fairly good approximations to the
systemic velocity and stellar velocity dispersion in all but very radio-loud
AGN~\citep{nelson96,sulentic00,greene05,crenshaw10}. In galaxies where both HI and HII lines have been observed,
the line profiles were found to match each other well
\citep[e.g.,][]{emonts05,holt08}. Matched profiles of HI and CO have been
found in NGC1266, which shows clear evidence for an AGN-driven molecular
wind \citep{alatalo11}.

Interpreted in light of this long legacy of optical observations, the
profiles of lines of ionized gas in SW022513 are a clear signature of gas
that is being stirred up and accelerated by the AGN, and which could be
outflowing \citep[see][for details]{nesvadba11b}. The detection of matching
CO and HII line profiles and the analogy with NGC\,1266 illustrate that this
wind may well have a molecular component.

Interestingly, LVG modeling of the CO(1-0) to CO(3-2) line ratios of NGC1266
suggests that the CO emission in the wing of NGC1266 is optically thin, akin
to what \citet{papadopoulos08,papadopoulos10} previously found in 3C293 at
$z$=0.04, another prime and relatively nearby example of an AGN-driven wind.
The gas becomes optically thin because CO molecules have a large range of
relative velocities, and consequently do not emit at strictly the same
wavelength. Given the immense width of the CO line in SW022513, we suspect
the same could be true here. Assuming that the gas conditions are broadly
similar to those in NGC1266 (which is not implausible if both are dominated
by the same basic astrophysical mechanism), we may overestimate the
molecular gas mass in the wing. We have requested multi-transition CO
observations of SW022513 to further investigate this question.

In this scenario, the CO emission could trace gas that is part of a
turbulent multiphase medium created by interactions between AGN and
ISM, similar to what is found in nearby radio galaxies
\citep[][]{nesvadba10a,ogle10}. In these environments, the molecular,
atomic, and ionized gas may be undergoing a permanent mass and energy
exchange, explaining very naturally why line profiles are very
similar. Molecular gas may even form in the stirred-up, turbulent,
post-shock gas \citep{guillard09}. This would circumvent the
difficulties of accelerating dense molecular gas with AGN-driven winds
\citep[e.g.,][]{sutherland07}.

NGC1266 may qualitatively be a good analog of SW022513, but quantitatively
it is not. The CO FWHM in SW022513 is much greater ($\sim$1000\,\kms\
compared to 350\,\kms), and the relative CO line flux in the broad component
of SW022513 appears much higher than that in NGC1266, about 89\%
(Table~\ref{co_params}) compared to $\sim$34\% \citep{alatalo11}. This could
have a number of reasons. First, the relative CO emissivity in the narrow
and broad components could be different in the two sources. In SW022513 we
observe CO(4-3), not CO(3-2), and would expect that this line is more
strongly boosted for optically thin relative to optically thick gas. Second,
while both studies suggest the outflow is driven by the mechanical (radio)
AGN luminosity, the mechanical energy injection rate in SW022513 is $>$30
times higher than in NGC1266~\citep[i.e.;
2$\times$10$^{44}$\,\ergs;~\citealt{nesvadba11b} $vs$
6$\times$10$^{42}$\,\ergs; ][]{alatalo11}. Both authors explicitly show that
the AGN provides sufficient energy to power the wind. Third, the covering
fraction of gas intercepted by the AGN may be greater in SW022513 than in
NGC1266, because the galaxy is richer in gas, or simply because of geometry.

A key question of AGN feedback in the context of galaxy evolution is: Will
the wind escape? Adopting the redshift of the narrow components of molecular
and ionized gas as the systemic redshift, we find that molecular line
emission extends to relative velocities on the order of $-$1000\,\kms\ from
systemic. Here we adopt the redshift corresponding to the bluest channels
where CO is detected with flux densities $\ge$1\,mJy to estimate the
terminal velocity. Terminal velocities are commonly adopted in wind studies
\citep[e.g.,][]{heckman00} to estimate the intrinsic outflow velocity,
irrespective of projection and other effects, which would only lower the
observed relative to the intrinsic velocity. This terminal velocity is on
the order of the escape velocities typically estimated for massive
galaxies~\citep[i.e., $\sim$900\,\kms;][]{vanzella10}, suggesting that at
least a fraction of this gas could potentially escape.

\subsection{Star-formation efficiency}\label{sfe}

The star-formation efficiency (SFE),  defined as the SFR per unit of
star-forming gas, i.e., SFR/M(H$_2$), can be derived from observables
(assuming the L(FIR)-SFR correlation) in terms of the continuum-to-CO line
luminosity ratio L(FIR)/L$^{\prime}_{CO}$. The CO luminosity provides a
measure of gas mass~\citep{solomon87}, and the FIR luminosity of the
SFR~\citep{kennicutt98a}. The L(FIR)/L$^{\prime}_{CO}$ ratio thus indicates
the efficiency with which the gas is converted into stars~\citep[a spatially
integrated version of the Schmidt-Kennicutt star-formation
law;][]{schmidt59,kennicutt98a}. The FIR luminosity, L(FIR), is derived as
(0.56$\pm$0.01)$\times$L(IR)~\citep{polletta08c}, where L(IR) is the
luminosity in the wavelength range 8--1000\,\micron, and L(FIR) the
luminosity in the 40--120\,\micron\ range. The estimated values for
SW022550, and SW022513 are, respectively, (2--11)$\times$10$^{12}$\,\lsun,
and (2--9)$\times$10$^{12}$\,\lsun~\citep{polletta08c}.

Previous studies find that the SFE increases with L(FIR), implying that star
formation is more efficient in ultra luminous systems, like submillimeter
galaxies, than in normal starforming galaxies~\citep{greve05,daddi10b}.
Powerful systems at high redshift, like SMGs, QSOs, radio galaxies (RGs),
and ULIRGs, exhibit average SFE values ranging from about 200 to
400~\lsun/(K\,km\,s$^{-1}$\,pc$^2$)~\citep[see~\citealt{iono09} and
\citealt{riechers11} for a recent compilation and
comparison]{greve05,solomon05,tacconi08,coppin08b,aravena08}. An order of
magnitude lower SFEs are instead observed in local spirals~\citep{leroy09}
and gas rich massive disk galaxies at $z$$\sim$1.5, i.e.
BzKs~\citep{daddi10b}.

\begin{figure}
\centering
\includegraphics[width=9cm]{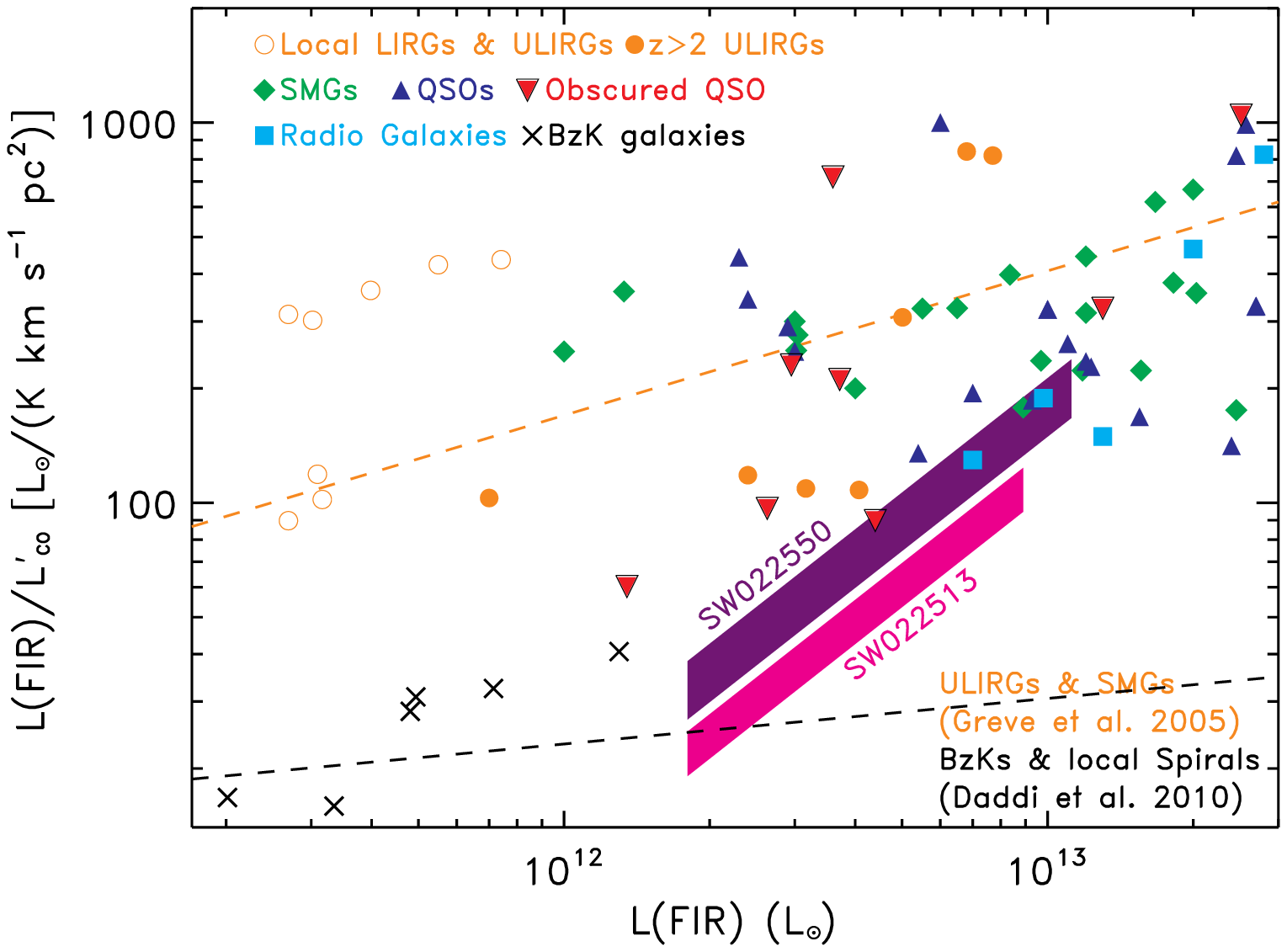}
\caption{Star-formation efficiencies (SFEs) given by the
L(FIR)/L$^{\prime}_{CO}$ ratio as a function of the FIR luminosity for local
LIRGs and ULIRGs~\citep[orange open circles;][]{solomon97,iono09},
$z\gtrsim$2 ULIRGs~\citep[orange full circles;][]{bothwell10,yan10},
SMGs~\citep[full green diamonds;][]{greve05,solomon05,tacconi08}, type I
QSOs~\citep[full blue triangles;][]{solomon05,coppin08b}, obscured QSOs
(upside-down full red triangles; see Table~\ref{qso2_tab}), radio
galaxies~\citep[full cyan squares;][]{solomon05}, and BzK
galaxies~\citep[black crosses;][]{daddi10b}. The dashed orange line is the
best-fit relation obtained from a sample of ULIRGs and SMGs
by~\citet{greve05}, and the black dashed line the relation obtained from
local spirals and BzKs by~\citet{daddi10b}. The two filled regions represent
the SFEs estimated for SW022550 (purple) and SW022513 (magenta). Note
that the FIR luminosities are affected by up to a factor of 2
uncertainties owing to the different methods applied to estimate them
in the various samples, and to the uncertainties associated
with the measured FIR/sub-mm/mm fluxes used in those estimates.}
\label{lco_lfir} 
\end{figure}

The range of SFE values derived for SW022550 and SW022513 are,
respectively, 35-194, and 24--108~\lsun/(K\,km\,s$^{-1}$\,pc$^2$) (see
Table~\ref{co_params}). The SFEs in our sources and in several types of
sources from the literature are shown in Figure~\ref{lco_lfir}. The sources
from the literature include type I QSOs, obscured AGNs 
(see~\S\ref{comparison}), RGs, SMGs, $z$$\sim$2 and local ULIRGs, and LIRGs
and BzK
galaxies~\citep{solomon05,coppin08b,aravena08,greve05,tacconi08,yan10,bothwell10,solomon97,iono09,daddi10b}.
We also show, as reference, the relationships between SFE and L(FIR) derived
from ULIRGs and SMGs~\citep{greve05}, and from local spirals and BzK
galaxies\footnote{Note that the SFE values reported for the BzK galaxies and
the relation from SFE and L(FIR) have been modified to be consistent with
the other values from the literature and with our estimates. In particular,
we derived the FIR luminosity from the SFR using the relationship L(FIR) =
0.56$\times$L(IR) =
0.56$\times$5.8$\times$10$^9$$\times$SFR~\citep{kennicutt98a,polletta08c}.
Note that this correction factor reduces their SFEs by a factor equal to
0.32. In \citet{daddi10b}, the mean SFE for BzKs is
84$\pm$12~\lsun/(K\,km\,s$^{-1}$\,pc$^2$), about three times higher than
reported in Figure~\ref{lco_lfir}.}~\citep{daddi10b}.

Our sources are among the powerful high-redshift systems with the lowest
SFEs, i.e., the highest estimates for our sources overlap with the lowest
third~\citep[$<$200~\lsun/(K\,km\,s$^{-1}$\,pc$^2$); ][]{solomon05} observed
in other powerful systems. With respect to the relation of ULIRGs and
SMG~\citep[Log(L$^{\prime}_{CO}$\,=\,(0.62$\pm$0.06)$\times$Log(L(FIR)+2.33$\pm$0.93;
][]{greve05} our sources' SFEs are about 1$\sigma$ lower.

To investigate whether low SFEs are common among obscured AGNs, we examine
the SFEs of the eight CO-detected obscured AGNs described in \S~\ref{comparison}
and listed in Table~\ref{qso2_tab}. Their gas masses and SFRs are consistent
with those observed in SMGs and type I QSOs. Note that exceptionally low, by
up to two orders of magnitude, gas masses and SFRs are observed in
IRAS\,F10214+4724. Because this source is magnified by a lensing object, it
was possible to detect it in CO even if it is intrinsically weaker than
those commonly detected with current facilities. The estimated SFEs in
obscured AGNs, also shown in Fig.~\ref{lco_lfir}, cover a broad range, from
60 to $>$1000\,\lsun/(K\,km\,s$^{-1}$\,pc$^2$), similar to that
observed in the other classes shown in Figure~\ref{lco_lfir}. Interestingly,
of all sources shown in Figure~\ref{lco_lfir}, those with the lowest SFEs
are obscured QSOs (see also Table~\ref{qso2_tab}), in particular three
sources have SFEs below 100\,\lsun/(K\,km\,s$^{-1}$\,pc$^2$). On the other
hand, obscured QSOs are also among the sources with the highest SFEs in
Figure~\ref{lco_lfir}. The obscured QSO sample is too small and incomplete
to be representative of the obscured AGN class. However, this analysis
indicates that this class can exhibit a wide range of SFEs, consistent with those
observed in other types of ULIRGs at high-$z$.

The relatively low SFEs in our sources and in some obscured AGNs
disfavor the hypothesis that the AGN might enhance star-formation activity
by compressing and cooling the molecular gas as suggested by some
models~\citep{begelman89,silk05,silk09}. If there were AGN-driven positive
feedback, the SFE in our sources would be higher compared to starburst
galaxies, e.g., SMGs, contrary to what is observed. The lower SFEs instead
point to a star-formation triggering mechanism that is less efficient than
what operates in sources with large SFEs. Thus, even if large amounts
of molecular gas are available, only a relatively small fraction is
converted into stars. Alternatively, their star formation might be shutting
down, perhaps as a consequence of the turbulence in the molecular gas, and
this might be a sign of AGN negative feedback, at least in SW022513. 

\subsection{Evolutionary phase}\label{evolution}

Many current evolutionary models postulate that high-$z$ powerful starburst
galaxies, obscured QSOs, and type I QSOs represent subsequent phases of a
system following a major merger~\citep[e.g.,][]{hopkins05b}. In order to
assess whether our sources follow this scheme, and, in such a case, which
phase they might represent, we estimate their molecular gas-to-stellar mass
ratio (M$_{gas}$/M$_{stellar}$), and black hole-to-stellar mass ratio
(M$_{BH}$/M$_{stellar}$) and compare them with those measured in
submillimeter galaxies and type I QSOs. A morphological analysis would also
be useful to locate our systems on this evolutionary sequence, but the
spatial resolution of the available images is not sufficient to carry it out.
Moreover, the host light may be diluted by the AGN at most wavelengths, or
highly extincted by dust.

Assuming the estimated stellar mass and associated
uncertainty~\citep[see Table~\ref{co_params}; ][]{polletta08c}, the gas
fractions, defined as the M$_{gas}$/M$_{stellar}$ ratio, are
8$\alpha$--36$\alpha$\% in SW022550, and 21$\alpha$--41$\alpha$\% in
SW022513, or 6--29\% and 17--33\%, respectively,
assuming $\alpha$\,=\,0.8\,\msun (K\,\kms\,pc$^2$)$^{-1}$
(see~\ref{co_intensity}). Note that these estimates have large systematic
uncertainties because they are based on CO(4-3) observations, and approximate
stellar masses. We may either miss a diffuse component, which could make
these fractions factors of $\sim$2 higher, or, alternatively, the CO line
emission may be optically thin and/or partially powered by shocks and/or
X-rays, which could lower the gas mass by up to factors of a
few~\citep{meijerink06,meijerink07,flower10}. These fractions are consistent
with the wide range of gas fractions estimated in SMGs~\citep[i.e.
$<$5$\alpha$--30$\alpha$\%;][]{tacconi08}, and lower than in high-$z$ disk
galaxies~\citep[30$\alpha$--50$\alpha$\%;][]{daddi10b}\footnote{Note that
the $\alpha$ factor derived for high-$z$ disk galaxies by~\citet{daddi10b}
is 3.6, compared to 0.8 which is commonly assumed for ULIRGs and adopted for
SW022550, and SW022513.}. Thus, our targets are as gas-rich as SMGs. The
lack of stellar mass estimates for type I QSOs and other obscured QSOs does
not allow us to estimate the gas fraction for these types of sources. In the
literature, the gas fraction in high-$z$ systems, like SMGs and type I QSOs,
is usually computed as the ratio between the gas and the dynamical mass, but
because the latter is highly uncertain owing the unknown system geometry, we
preferred to consider only the gas and stellar masses for this comparison.

Another indication of the evolutionary stage is provided by the comparison
between the BH and stellar mass. As has been argued recently by a number of
authors~\citep{dimatteo08,alexander08,coppin08b,nesvadba11a,sarria10}, the
position of a high-redshift galaxy on the BH-bulge mass relationship can be
used to infer whether it is in an early or advanced evolutionary stage.
Lower limits to the BH masses of our sources, obtained assuming that they
are radiating at the Eddington limit, are $>$7.4$\times$10$^8$\,\msun, and
$>$3.7$\times$10$^8$\,\msun, for SW022550 and SW022513,
respectively~\citep{polletta08c}. A more robust (virial) BH mass estimate
can be obtained for SW022550 from the width of the \civ\ line and the
optical luminosity. Namely, we base our estimate on the BH mass estimator
of~\citet{vestergaard06}, Log(M$_{\rm BH}$)= Log[FWHM$_{3}^2$
($\lambda$L$_{\lambda,44}$)$^{0.53}$]+6.66, where the FWHM$_3$ is the \civ\ FWHM
in units of 10$^3$\,\kms, and $\lambda$L$_{\lambda,44}$ is the luminosity at
1350\AA\ in units of 10$^{44}$\,\ergs. This relationship is valid for type I
QSOs, whereas SW022550 is obscured. Because obscuration does not permit a
direct measurement of the AGN UV luminosity, we derive the AGN luminosity at
1350\AA\ indirectly from the bolometric AGN luminosity, using the bolometric
correction of~\citet{elvis94a}, L$_{bol}$/L$_{UV}$$\sim$6\footnote{We note
that the bolometric luminosity is the quantity that black-hole mass scales
with intrinsically, and is approximated by the monochromatic UV luminosity
for mere practical reasons.}

The estimated AGN bolometric luminosity is
10$^{47}$\,\ergs~\citep{polletta08c}, which corresponds to a luminosity of
1.6$\times$10$^{46}$\,\ergs\ at 1350\AA. The \civ\ line width,
FWHM(\civ)$=$(3690$\pm$100)\,\kms~\citep{polletta08c}, implies M$_{\rm
BH}$\,=\,(9.2$\pm$0.5)$\times$10$^8$\,\msun, consistent with the estimate
in~\citet{polletta08c}. Because a BH with such a mass corresponds to
an Eddington luminosity of 1.2$\times$10$^{47}$\,\ergs, this estimate
implies that SW022550 is radiating at $\sim$85\% of the Eddington-limit.

Assuming a BH mass of 9$\times$10$^8$\,\msun\ for SW022550, and
$\geq$4$\times$10$^8$\,\msun\ for SW022513, the M$_{\rm BH}$/M$_{star}$
ratios of our sources are 0.001--0.006 for SW022550 and
$\gtrsim$0.001--0.002 for SW022513. These are consistent with the local
relationship~\citep[$<$M$_{\rm
BH}$/M$_{bulge}$$>$$\sim$0.0014;][]{haring04,marconi03}, and also with the
ratios measured in type I QSOs~\citep[0.007$\pm$0.003;][]{peng06b}, but
larger than those estimated in SMGs~\citep[0.0001--0.0008;][]{alexander08}.
Several recent models of galaxy evolution postulate that only relatively
evolved supermassive black holes may be able to affect their surrounding
gas~\citep{churazov05,merloni08,fanidakis11}. As argued by,
e.g.,~\citet{coppin08b,alexander08,nesvadba11a}, this would imply that the
AGN and their host galaxies would approach the local black-hole bulge
scaling relationship near the end of their active growth period. For many
high-redshift quasars, this seems not to be the
case~\citep[e.g.,][]{walter04,peng06b}, perhaps indicating that they have
not yet reached this stage. On the other hand, the position of our sources
on the local M$_{\rm BH}$--M$_{star}$ relationship suggests that both
stellar and black hole growth might have reached the end of their most
active phase, and that our sources are thus at a more advanced evolutionary
stage with respect to type I QSOs and SMGs. This could be a consequence of
AGN feedback if a sufficiently large fraction of the gas is heated and
blown out of the galaxy. This result is consistent with the idea, suggested
by their low star-formation efficiency (see \S~\ref{sfe}), that our sources,
and perhaps other obscured AGNs with low SFEs, might be on the verge of
turning off.

\section{Summary and conclusions}

We presented the properties of the molecular gas and the possible role of
powerful AGN activity on this gas in two highly luminous obscured QSOs at
$z$$>$3.4.

Our study was based on observations of the CO(4--3) line carried out with the
PdBI. The luminosity and profile of the CO line indicate that both systems
contain large amounts of molecular gas
(M$_{gas}$$\gtrsim$4$\times$10$^{10}$\,\msun) corresponding to $<$30\% of
their stellar mass. The CO lines are among the broadest CO lines ever
observed ($\sim$800--1000\,\kms). A double-peaked line is observed in
SW022550, and a blueshifted broad profile in SW022513. Various scenarios
were discussed to explain the observed line profiles and gas kinematics: a
rotating disk, a merger, and an AGN-driven outflow.

The CO properties in SW022550 indicate that the molecular gas might be
distributed in a highly inclined, almost edge-on, rotating disk or be
associated with a merger. The CO emission appears extended on scales of
$\sim$16\,kpc, consistent with both scenarios. The disk inclination is constrained
by the presence of two peaks in the line profile and by the high estimates
of rotational velocities and dynamical masses. A merger can also
explain the CO line kinematics, even if there is no evidence of multiple
sources in the available optical and infrared images. A comparison with the
rest-frame optical spectrum of the source~\citep{nesvadba11b} suggests that
the molecular gas is not significantly affected by the presence of the AGN.

For SW022513, the disk and merger hypotheses seem less convincing. Based on
an analysis of the rest-frame optical line emission in SW022513, we suspect
that parts of the molecular gas could be stirred up by the AGN and perhaps
be part of an AGN-driven outflow. This is further supported by a detailed
comparison with NGC1266, the clearest example of an AGN-driven molecular
wind in a nearby galaxy. In this scenario, most of the line flux would be
produced in a very turbulent and potentially outflowing wind component. The
wind hypothesis would very naturally explain the similarity of the CO and
H$\beta$ line profiles.

A comparison of our objects' molecular gas properties with those of other
powerful high-redshifts systems, i.e., submillimeter galaxies, type I QSOs,
RGs, and $z$$\gtrsim$2 ULIRGs, shows that they are characterized by
relatively lower star-formation efficiencies, defined as the
L(FIR)/L$^{\prime}_{CO}$ luminosity ratio. Similarly low SFEs are also found
in other obscured AGNs from the literature, but they are not typical of
this class of objects. Another difference between our sources and SMGs and
type I QSOs it is that they already seem to lie near the local BH-bulge mass
relation. These results suggest that our sources, and perhaps other
obscured QSOs, might be near the end of their most active growing phase and
on the verge of turning off and transitioning into passively evolving
massive galaxies. Their low star-formation efficiencies might be due to the
perturbation caused by the AGN on the molecular gas.

Our observations indicate that the AGN plays a major role in perturbing the
molecular gas in SW022513. This source might thus provide a laboratory to
test and constrain evolution models and to study the formation of massive
galaxies at critical early epochs. It could also be used as reference to
search for similar sources and carry out more targeted searches for
molecular gas outflows.


\begin{acknowledgements} 
 We thank the referee for a careful reading of the manuscript and helpful
 comments and suggestions that improved both the content and the clarity of
 the paper. MP is grateful to the Institut d'Astrophysique de Paris for its
 hospitality during her visit there. NPHN wishes to thank P.  Papadopoulos
 and M. D. Lehnert for interesting discussions about various aspects of this
 work. This work is based on observations carried out with the IRAM Plateau
 de Bure Interferometer.  IRAM is supported by INSU/CNRS (France), MPG
 (Germany) and IGN (Spain). We thank the staff of the IRAM Observatory for
 their support of this program.  MP acknowledges financial contribution from
 contract ASI-INAF I/016/07/0.
\end{acknowledgements}

\end{document}